\documentclass[sn-mathphys,Numbered]{sn-jnl}
\usepackage{graphicx}
\usepackage{amsmath,amssymb,amsfonts}
\usepackage[title]{appendix}
\usepackage{xcolor}
\usepackage{manyfoot}
\usepackage{subcaption}
\usepackage{hyperref}

\begin{document}

\title[Learning to Reconstruct: A Differentiable Approach to Muon Tracking at the LHC]{Learning to Reconstruct: A Differentiable Approach to Muon Tracking at the LHC}

\author[1]{Andrea Coccaro}
\author[2,3]{Francesco Armando Di Bello}
\author*[1,4]{Lucrezia Rambelli}
\email{lucrezia.rambelli@ge.infn.it}
\author[5]{Stefano Rosati}
\author[1,4]{Carlo Schiavi}
\affil[1]{INFN, Sezione di Genova, 16146 Genova, Italy}
\affil[2]{Department of Physics, Università di Pisa, 56127 Pisa, Italy}
\affil[3]{INFN, Sezione di Pisa, 56127 Pisa, Italy}
\affil[4]{Department of Physics, Università di Genova, 16146 Genova, Italy}
\affil[5]{INFN, Sezione di Roma, 00185 Rome, Italy}

\abstract{
Reconstructing the trajectories of charged particles in high-energy collisions requires high precision to ensure reliable event reconstruction and accurate downstream physics analyses. In particular, both precise hit selection and transverse momentum estimation are essential to improve the overall resolution of reconstructed physics observables. Enhanced momentum resolution also enables more efficient trigger threshold settings, leading to more effective data selection within the given data acquisition constraints.
In this paper, we introduce a novel end-to-end tracking approach that employs the differentiable programming paradigm to incorporate physics priors directly into a machine learning model. This results in an optimized pipeline capable of simultaneously reconstructing tracks and accurately determining their transverse momenta.
The model combines a graph attention network with differentiable clustering and fitting routines, and is trained using a composite loss that, due to its differentiable design, allows physical constraints to be back-propagated effectively through both the neural network and the fitting procedures. 
This proof of concept shows that introducing differentiable connections within the reconstruction process improves overall performance compared to an equivalent factorized and more standard-like approach, highlighting the potential of integrating physics information through differentiable programming.}


\maketitle

\section{Introduction}\label{sec:intro}
Track reconstruction is a fundamental task in particle-physics experiments, as it provides key observables -- particle trajectories, momenta and vertex information -- essential for physics analyses. Traditional tracking approaches in High Energy Physics rely on a factorized pipeline, where pattern recognition and track fitting are treated and optimized as separate steps. With the forthcoming increase in luminosity and data complexity at the LHC~\cite{HL-LHC}, developing optimized methods for tracking has become an important research direction~\cite{GNN-Review,Tracking-Service}, also considering the important detector upgrades of the ATLAS and CMS collaborations~\cite{ATLAS-Paper,CMS-Paper}.

Furthermore, recent interesting developments have focused on integrating physics information into machine learning models to improve their accuracy, with differentiable programming techniques proving to be highly effective for implementing these \mbox{priors~\cite{differentiable-tracking,INFERNO,neos,endtoend,tracking-belle,vertex}}.

In this context, we designed and implemented a novel machine-learning based model which represents an end-to-end, differentiable approach to track reconstruction. To demonstrate the effectiveness of the proposed technique, we focus, as a concrete use case, on single muon track reconstruction of a typical LHC experiment~\cite{ATLAS-muon-reco,CMS-muon-reco}, restricting the problem on the 2D plane containing the particle's trajectory.

Traditional tracking methods typically rely on a sequence of separate, factorized steps. In particular using as inputs the hits registered by the detector by the traversing particles a clustering procedure and, subsequently, a fit procedure is applied  to accurately determine the particle's transverse momentum ($p_\text{T}$).

Unlike such traditional methods, the proposed model, thanks to the definition of differentiable blocks which allows to have a physics-informed weight backpropagation, directly maps raw detector hits to the particle’s transverse momentum within a single integrated pipeline. The model is based on a Graph Attention Network (GAT)~\cite{GAT}, representing detector hits as graph nodes and exploiting their spatial correlations to identify those associated with a single muon trajectory. Clustering and fitting operations are embedded as differentiable modules, allowing the entire process, from hit classification to momentum estimation, to be jointly optimized through backpropagation. This design makes the reconstruction inherently physics-informed and capable of learning the track curvature directly from the data.
The model is developed in a JAX-based environment~\cite{JAX}, which provides automatic differentiation, just-in-time compilation, and vectorization, ensuring both flexibility and computational efficiency. Applied to a toy simulation based on the ATLAS muon-spectrometer geometry and noise, the model demonstrates promising results in both hit classification and transverse momentum estimation and outperforms a sequential baseline approach. These achievements show the potential of differentiable, end-to-end techniques to become an alternative to traditional tracking algorithms.

\section{Detector simulation}\label{sec:simulation}
In order to develop the benchmark models considered in this study, a MC simulation of a tracking detector has been designed. 

\begin{figure}[t!]
\centering
\includegraphics[width=1\linewidth]{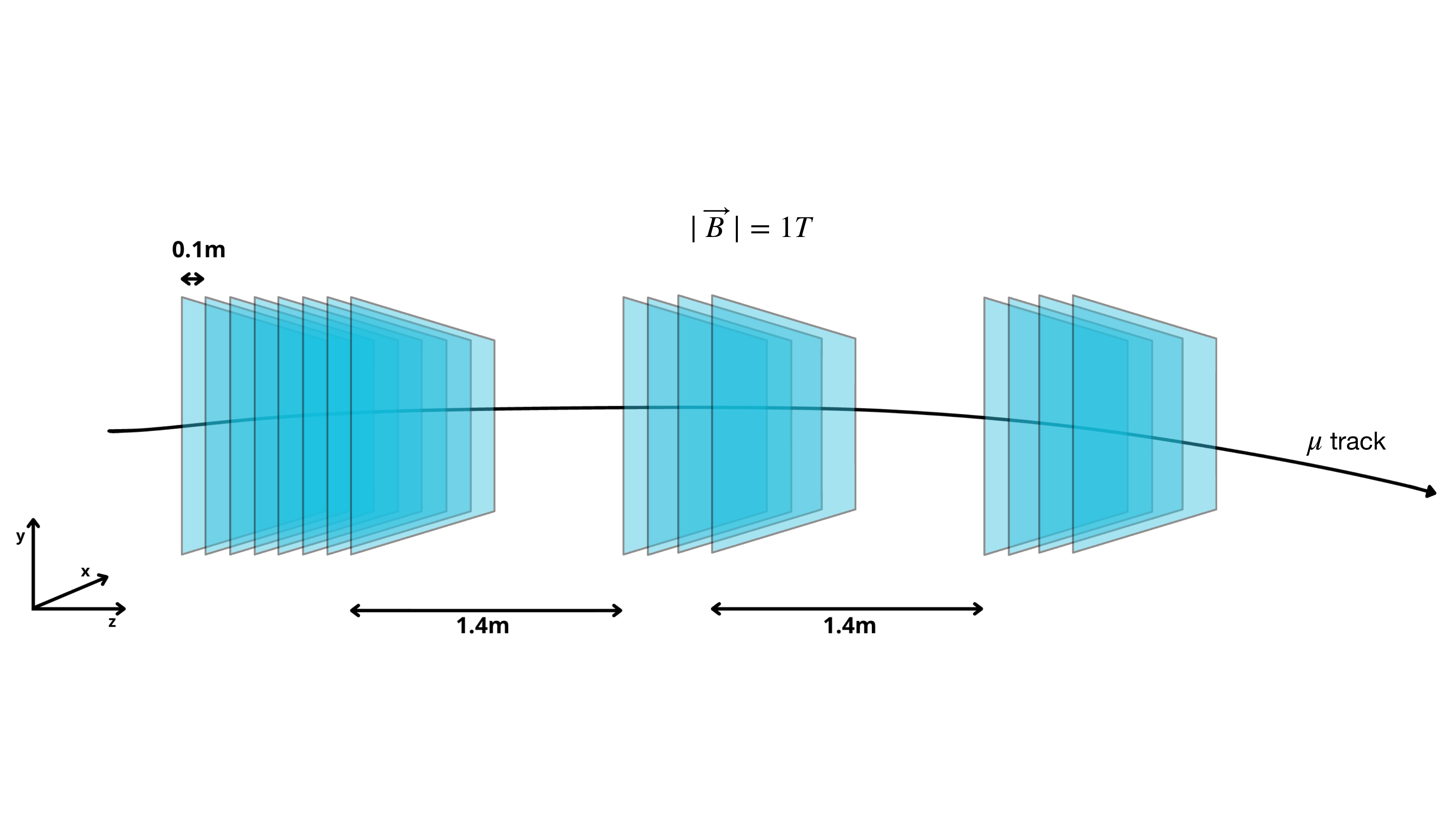}
\caption{Layout of the detector geometry simulated in the toy model}
\label{fig:detector}
\begin{subfigure}{0.49\textwidth}
\includegraphics[width=\linewidth]{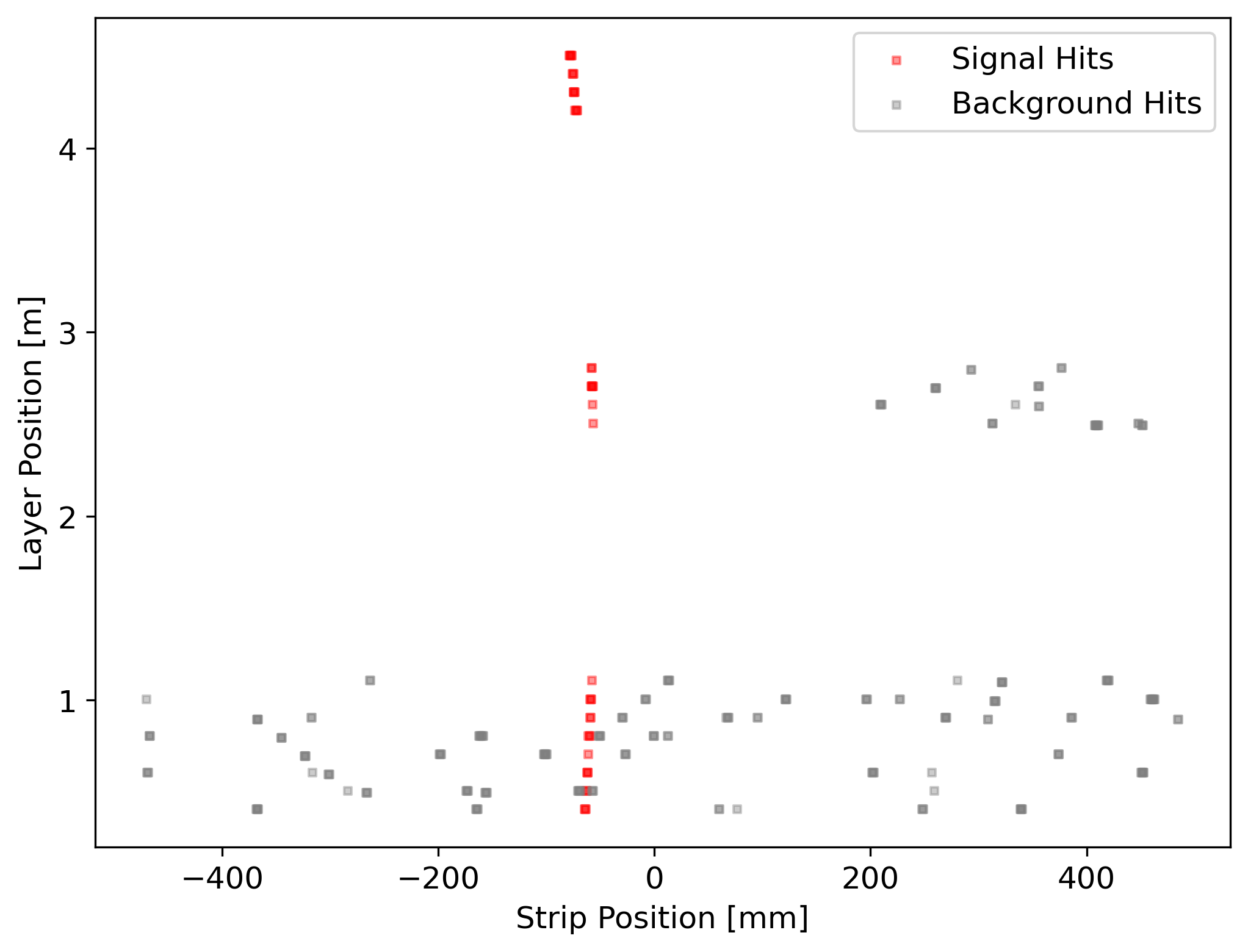}
\caption{}
\end{subfigure}
%\hfill
\begin{subfigure}{0.49\textwidth}
\includegraphics[width=\linewidth]{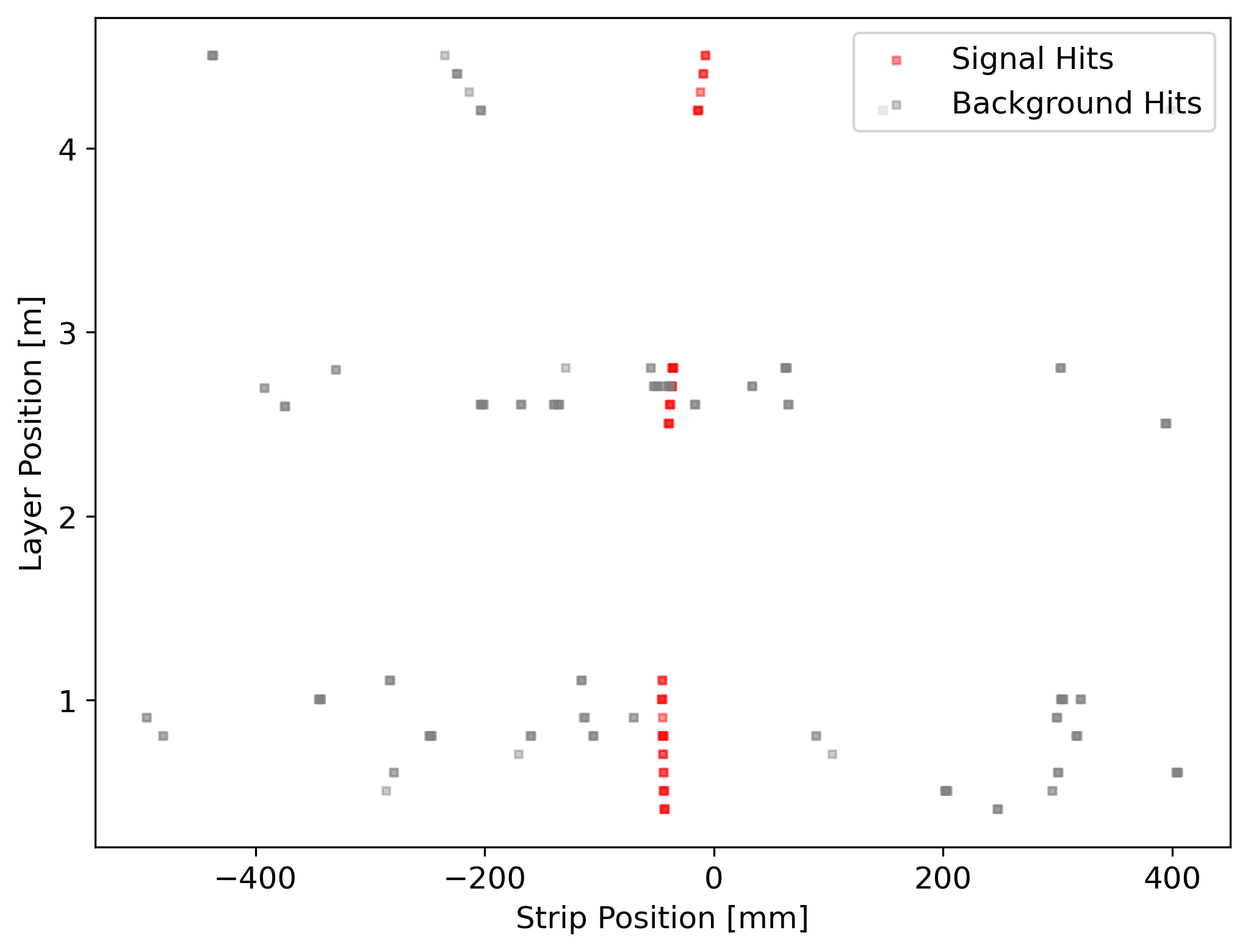}
\caption{}
\end{subfigure}
\caption{Two examples of event displays, both showing the track originating from the muon particle and additional activity related to both detector noise and secondary tracks.}
\label{fig:inputs}
\end{figure}

The detector layout, depicted in Figure~\ref{fig:detector},  is composed by 16 tracking layers, divided into three stations. The first station has 8 layers each, while the other two stations have 4 layers each. The detector is immersed in a 1\,T uniform magnetic field directed along the $x$-axis, as depicted in the figure. The simulation of this layout has been realized with \textsc{Geant4}~\cite{Geant4}, assuming that each detector layer is composed by two printed circuit boards with a 1\,cm thickness, interleaved with a sensitive gap of 5\,mm and a strip pitch of 0.5\,mm. This layout is inspired by micro-pattern gaseous detectors used in the LHC experiments, such as the MicroMegas~\cite{Micro-megas} or the GEM~\cite{GEM} detectors, and in particular by the design of the Phase-I ATLAS muon upgrade~\cite{ATLAS-Micro-megas}.

The detector response is simulated by extrapolating the particle trajectories to the readout planes, and getting the hit position on each plane. A cluster of firing strips is built around the position of the hit, with an average cluster size of 4 strips. The charge per strip is generated according to a gaussian distribution, centered in the hit position with a sigma corresponding to 2/3 of the strip pitch. This ensures that a charge always very close to the full cluster charge is contained within the cluster. If two or more hits generate overlapping clusters, the charge contributions are summed over all clusters, for each readout channel in the overlap region.

The simulated sample is composed of single muons events with a transverse momentum in the range $[20,50]$ GeV. Examples of event displays, with hits from signal tracks and from both noise or additional scattering tracks, are depicted in Figure~\ref{fig:inputs}.

\section{Model}\label{sec:model}
Tracking, the task of reconstructing particle candidates from a set of detector hits $x$, is composed of several algorithmic stages. 
Considering as input variables the geometrical coordinates in the $yz-$plane of each detector hit, the reconstructed transverse momentum of the particle  $\hat{p}_\phi$ is obtained as

\begin{equation*}
\hat{p}_\phi = f_{\chi^2\text{-min}} \circ f_{\text{clustering}} \circ f_{\phi}(x) \,.
\end{equation*}

As outlined in the above equation, the entire processing chain is differentiable, allowing end-to-end optimization of $\hat{p}_\phi$ through gradient-based learning.

The function $f_{\phi}$ represents the pattern recognition module, implemented as a GAT with trainable parameters $\phi$. It receives detector signals as input and outputs a set of weights $w_i$, corresponding to the probability that the $i$-th hit was produced by a signal muon. More details on the actual architecture of the GAT are presented in Appendix~\ref{app:model}.

Using as input the predicted weights for each graph node and the original features representing the spatial coordinates, the clustering module, $f_{\text{clustering}}$, computes the muon position $\bar{y}_j$ in each detector layer $j$ as a weighted average of hit positions:
\begin{equation*}
\bar{y}_j = \frac{\sum_{i \in j} w_i \, y_i}{\sum_{i \in j} w_i}\,,
\end{equation*}
where $\bar{y}_j$ denotes the reconstructed muon position in layer $j$.

\begin{figure}[t!]
\hspace{-1cm}
\includegraphics[width=1.2\linewidth]{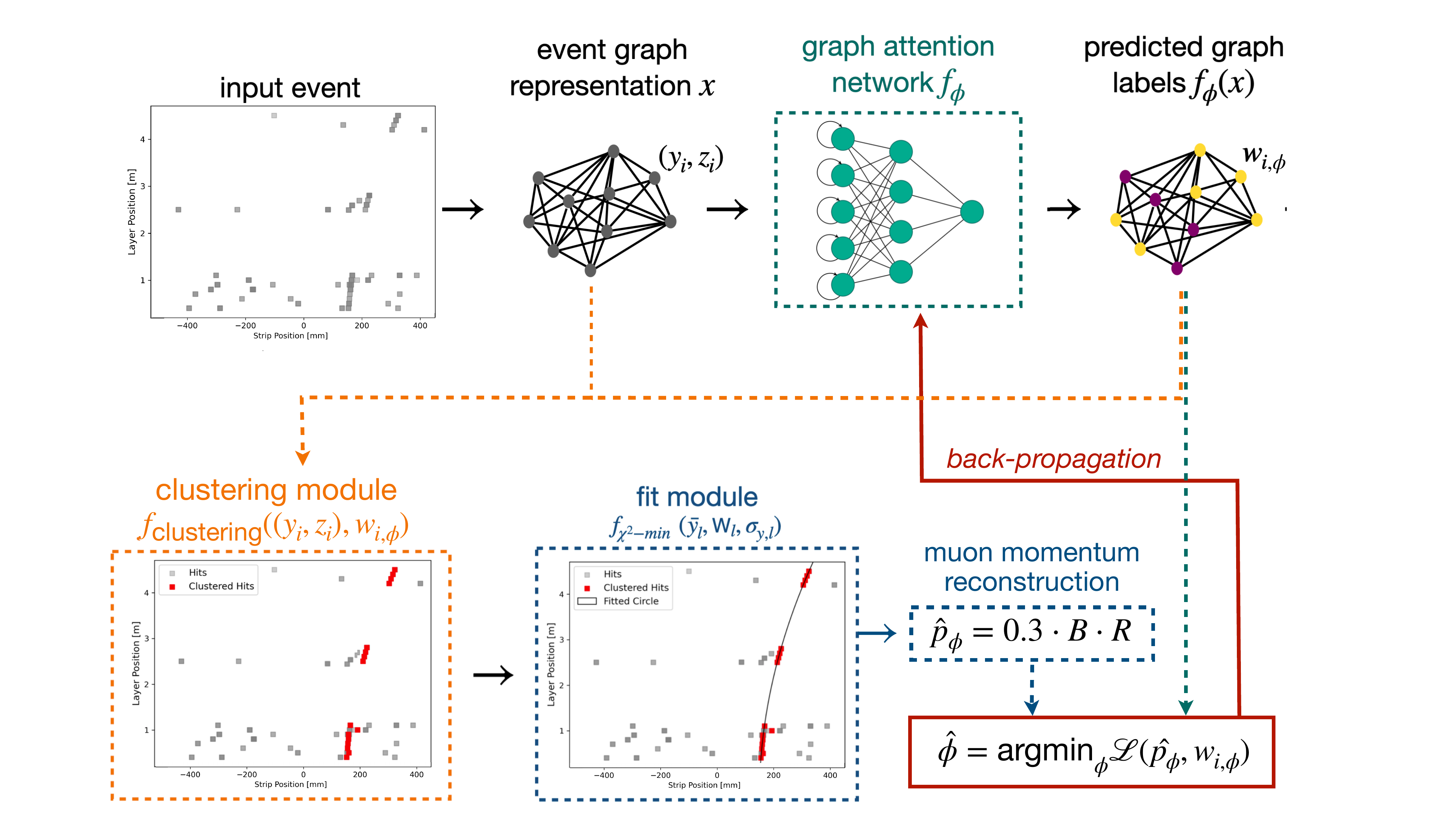}
\caption{Schematic representation of the end-to-end model developed for track reconstruction. Both the clustering and fitting routines depend on the GAT learnable parameters through the predicted node labels $w_{i,\phi}$. The regression of the particle's $p_\text{T}$ is back-propagated to the pattern recognition module as visualized in the figure and described in more details within the text.}
\label{fig:model_scheme}
\end{figure}

The RMS of the detector hits associated with the muon in that layer is used as the positional uncertainty $\sigma_{\bar{y_j}}$. An additional weight $\textbf{W}_j$ is then associated to each cluster and defined as the sum of all the predicted weights $w_i$ for the hits in the given layer $j$ used for the cluster position definition.

The $\chi^2$ minimization stage $f_{\chi^2\text{-min}}$ implements the fitting procedure using as input the reconstructed positions $\bar{y}_j$ and a matrix inversion procedure based on the Kasa method~\cite{KASA}.  Layer-wise weights, defined as $\textbf{W}_j/\sigma_{\bar{y}_j}^2$, are incorporated in each matrix element. The corresponding output is the reconstructed muon momentum $\hat{p}_\phi$ through the relation $\hat{p}_\phi = 0.3 \, B \, r$ and with $r$ the fitted radius.

The network trainable parameters are determined by minimizing the following composite loss function:
\begin{align*}
\hat{\phi} &= \arg \min_{\phi} \, \mathcal{L}(\hat{p}_{\phi}, w_{i,\phi}) = \\
           & =  \arg \min_{\phi} \sum_{\text{events}} \left[
            \alpha \frac{(\hat{p}_{\phi} - p)^2}{p^2}
           + \frac{1}{N_{\text{hits}}} \sum_{i=1}^{N_{\text{hits}}} \text{BCE}(w_{i,\phi}, y_{i,\text{truth}})
           \right] \, .
\end{align*}
This loss function consists of two distinct components, targeting both hits classification and momentum regression tasks. Here, the momentum $p$ in the loss term reflects the true $p_T$ value for each event, while $y_{i,\text{truth}}$ designates the truth labels corresponding to each individual node within the graph.
The hyperparameter $\alpha$ balances the regression and classification components and it is empirically set to 0.5. During the early training epochs, $\alpha$ is set to 0 so that only the binary cross-entropy (BCE) term is initially considered.

This two-stage approach was observed to yield a more rapid convergence of the training process allowing a comparison of the baseline and end-to-end models without any change on the actual model architecture. An interesting venue is the direct $p_T$ regression from the pooled graph-level embedding which was found to provide a less stable convergence in our trainings. This would require, for instance, operating on cluster-level embeddings and incorporating attention-based pooling. Such changes would increase the architectural complexity and are considered beyond the scope of the current study.

The model is trained on a dataset of $10^4$ events, while evaluation is performed on an independent test sample of 9k events. Optimization was performed using Adam~\cite{Adam}, with an OPTAX scheduler~\cite{Optax} that gradually decays the learning rate.

\section{Results}\label{sec:results}
The performance of the proposed end-to-end model is evaluated using two complementary criteria: hit-classification accuracy and reconstructed $p_\text{T}$ resolution. These metrics assess the model’s ability to discriminate signal hits from background noise and to accurately reconstruct particle kinematics.

The performance is compared to what is obtained by another model with the same architecture but trained solely for the node classification task, hence by setting $\alpha = 0$ in the composite loss function. This approach mimics the standard sequential workflow for tracking, in which hit classification and trajectory reconstruction are factorized and executed independently. In particular the $p_\text{T}$ reconstruction is done applying the clustering and circle-fitting steps using the same definitions and weighting schemes used for the main model, but outside of the end-to-end pipeline. Throughout the subsequent discussion and figures, these two models are referred to as the end-to-end and baseline models.

\begin{figure}[b!]
\centering
\includegraphics[width=0.8\linewidth]{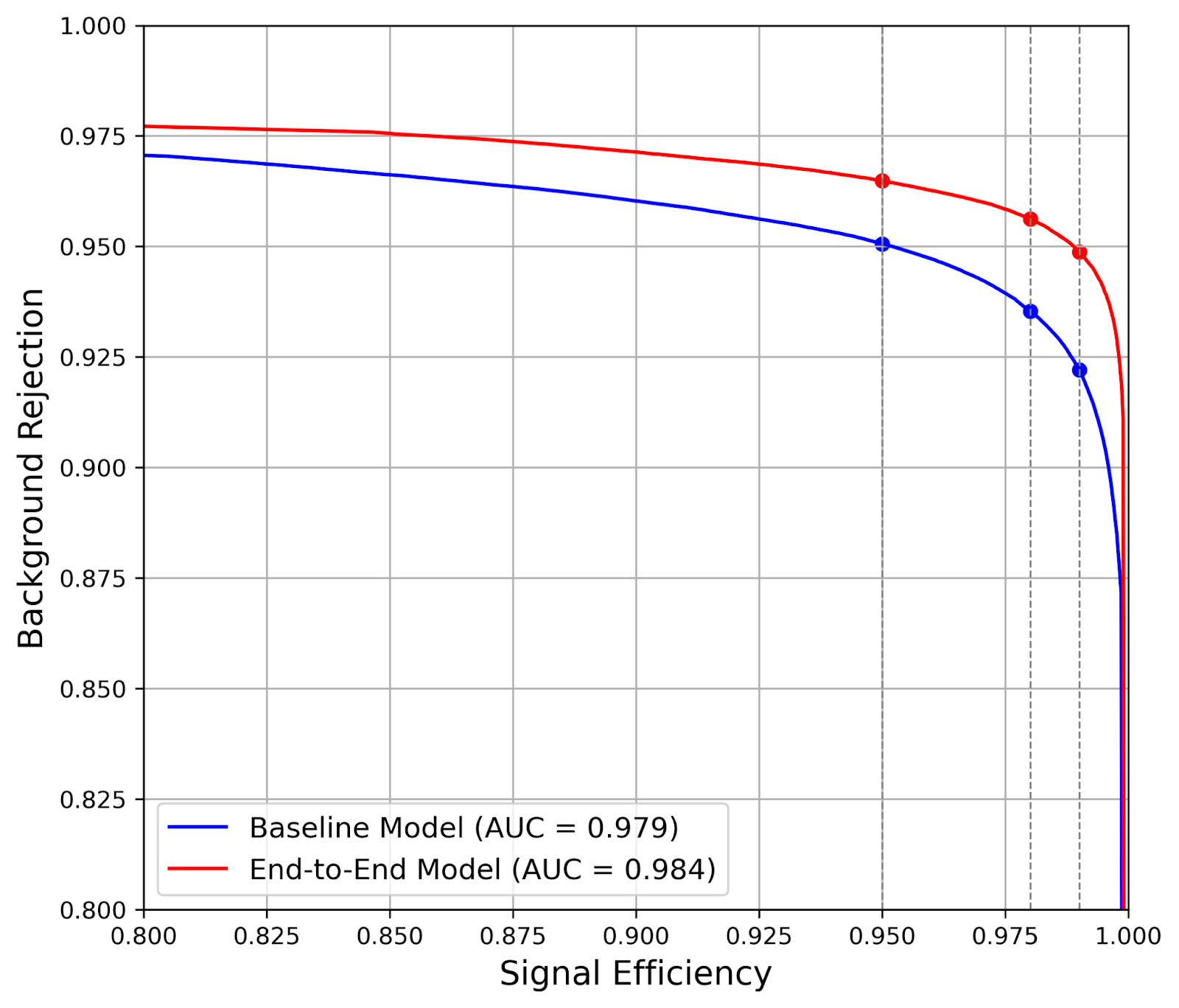}
\caption{ROC distributions for the hit classification task for both the baseline and the end-to-end models for tracking. The end-to-end model directly incorporates physics priors and perform simultaneously both the hit classification and the transverse-momentum regression. The performance clearly shows how the end-to-end approach boosts the accuracy of the hit classification task.}
\label{fig:ROC}
\end{figure}

For the classification task, performance is evaluated by constructing score histograms for true signal and background hits, and by generating a receiver operating characteristic (ROC) curve through threshold scanning over the predicted scores. The resulting plot, depicted in Figure~\ref{fig:ROC}, shows how including the $p_\text{T}$ loss term improves classification performance and demonstrates that the differentiable end-to-end coupling of the physics-driven $p_\text{T}$ objective benefits the node-level predictions. The AUC values are displayed in the plot and correspond to 0.979 and 0.984 for the baseline and the end-to-end model, respectively. By defining operating points corresponding to a signal selection efficiency of 95\%, 98\% or 99\%, the corresponding background rejections for the baseline and end-to-end models are evaluated in being 95\% and 97\%, 94\% and 96\%, or 92\% and 95\%.

A complementary study was performed to assess the spatial resolution of the reconstructed cluster positions. The analysis examines the residual distributions between the reconstructed cluster coordinates and the truth muon impact point on each detector layer.
Residual distributions were evaluated for three configurations: the idealized truth-weighted clustering, the baseline model, and the proposed end-to-end model. Using a Gaussian fit to the truth distribution to define the detector’s core resolution window, we measure the fraction of clusters with residuals within this region. The end-to-end model increases this fraction by 12.7$\%$ relative to the baseline, demonstrating a clear improvement in spatial reconstruction performance.

\begin{figure}[t!]
\centering
\includegraphics[width=0.6\linewidth]{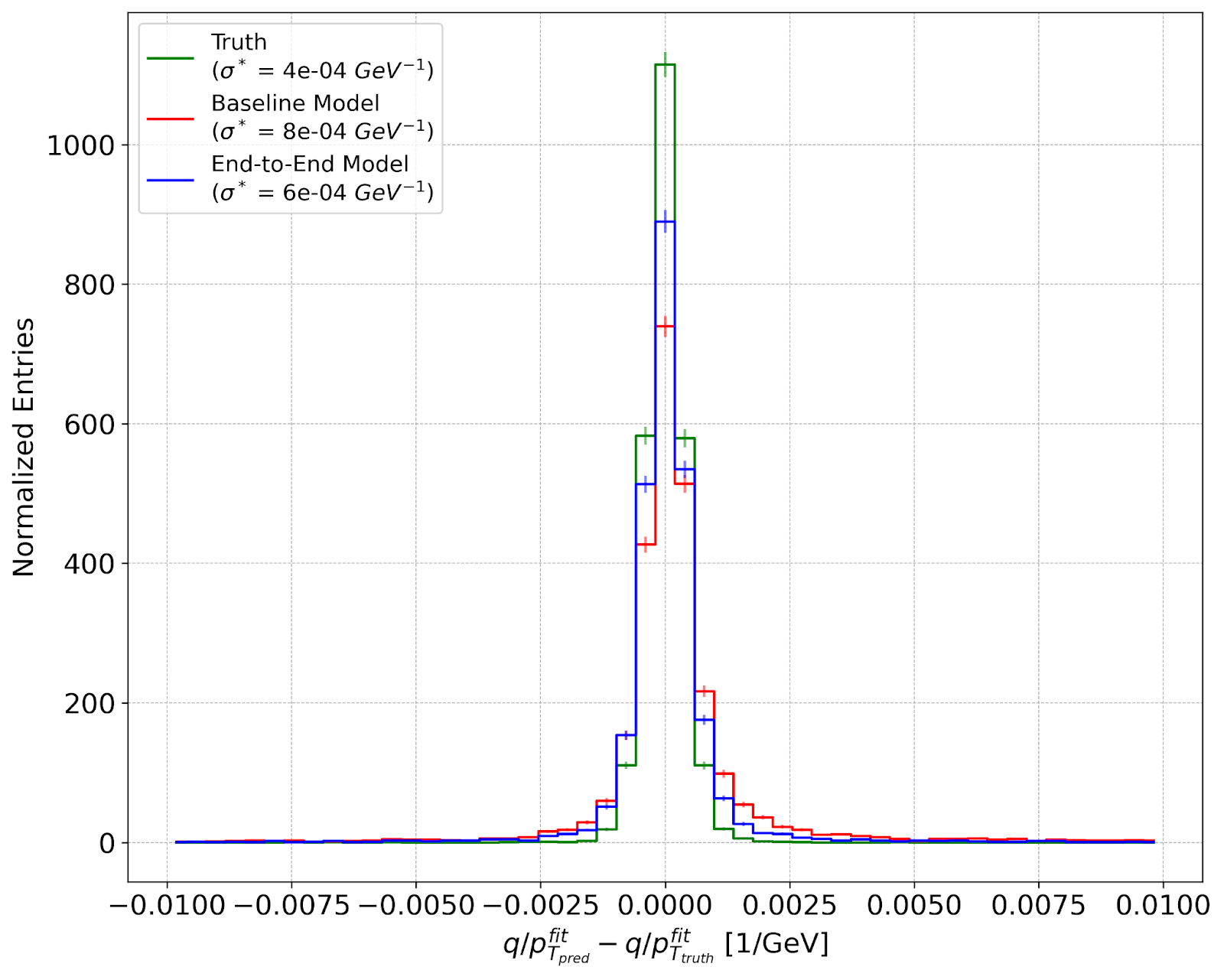}
\caption{Normalized $q/p_\text{T}$ residual distributions, with $q$ being the charge of the reconstructed muon. The three distributions correspond to using the truth labels as input weights in the clustering step, using the baseline model predictions and those from the proposed end-to-end model. The end-to-end model demonstrates improved accuracy in $p_\text{T}$ regression, approaching the limit of an ideal detector.}
\label{fig:1_pt_res}
\end{figure}
\begin{figure}[t!]
\centering
\includegraphics[width=0.8\linewidth]{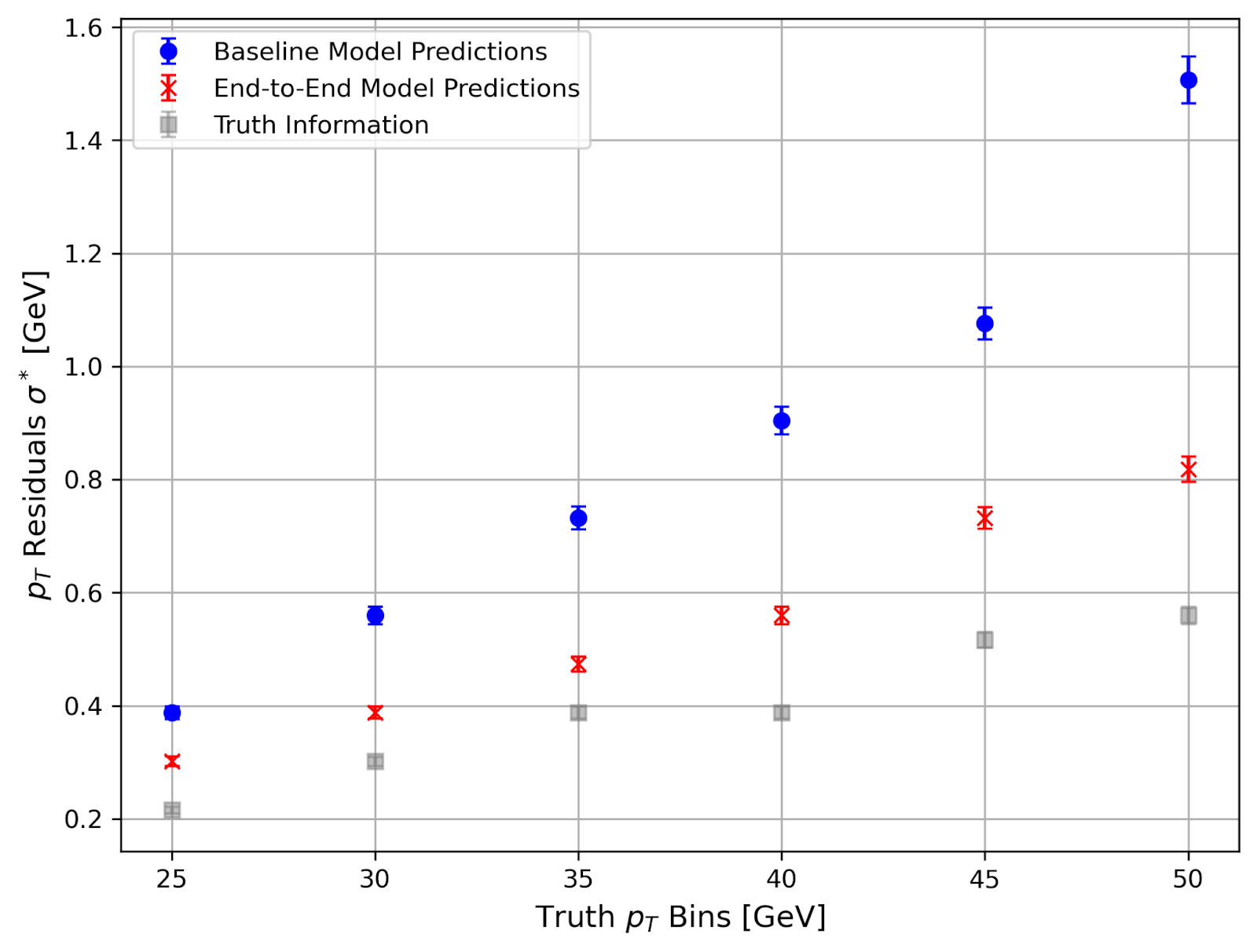}
\caption{$p_\text{T}$ residual distributions as a function of the truth $p_\text{T}$ values}
\label{fig:pt_res}
\end{figure}

In a similar fashion, the reconstructed $p_\text{T}$ accuracy is compared across these three different methodologies for assessing the impact of the end-to-end training approach. The reconstructed $p_\text{T}$ is evaluated by analyzing the $q/p_\text{T}$ residual distributions, with $q$ being the muon charge. Figure~\ref{fig:1_pt_res} shows these distributions for the three approaches. Following the procedure outlined in Reference~\cite{pT-study}, a width $\sigma^*$, defined as half the interval, centered around the median of the distribution and that contains $68\%$ of events, is calculated. 
The $\sigma^*$ distributions for the three cases are displayed in Figure~\ref{fig:1_pt_res}, which demonstrates a clear improvement in $p_\text{T}$ resolution for the end-to-end model relative to the sequential baseline.

We further study the residuals in the reconstructed $p_\text{T}$ and in 5-GeV bins across the 20–50 GeV range. In this case the truth $p_\text{T}$ value associated to each muon in the sample is used as a comparison, and the binned $\sigma^*$ vs. $p_\text{T}$ trend is computed again with using the truth labels as weights in the procedure and with the predictions from both the baseline and the end-to-end models. As clearly visible in Figure \ref{fig:pt_res}, across all bins the end-to-end approach yields systematically smaller $\sigma^*$ values, demonstrating improved resolution from incorporating the differentiable $p_\text{T}$ term in the training.

\section{Conclusions}\label{sec:conclusions}
In this work, we introduced a novel end-to-end tracking approach that incorporates differentiable programming techniques to embed physics information directly into the training of a machine-learning model.

The proposed model demonstrates a notable performance enhancement with respect to a more traditional approach where hit classification and fitting procedures are two sub-sequent and factorized steps of the reconstruction. Specifically, the ROC curve in Figure \ref{fig:ROC} shows how the end-to-end training yields a consistently higher background rejection at equivalent signal efficiencies for classifying detector hits. This demonstrates that integrating a physics-aware contribution within the loss function, particularly for regressing the muon $p_\text{T}$, enhances the model discriminative capabilities beyond what classification alone can achieve. This is made possible by the differentiable formulation of clustering and fitting functions used for the $p_\text{T}$ computation.

Overall, the combined evidence from ROC metrics and residual distributions consistently supports the advantages of the end-to-end, differentiable design. By embedding physics constraints into the training process, the end-to-end approach delivers improved hit classification accuracy and produces reconstructed clusters that more faithfully reflect true particle paths.

An important direction for future work is to assess the robustness and scalability of the proposed approach in more realistic detector conditions, such as non‑uniform magnetic field, time‑dependent inefficiencies in signal collection due to dead or noisy channels, and more complex hit topologies arising from overlapping particles. Incorporating these effects into the training and evaluation pipeline will be important for understanding how the differentiable components behave under imperfect detector responses and whether additional regularization or architecture optimizations are required. Extending the method to full detector simulations, and ultimately to experimental data, will also allow to study its performance in the presence of systematic uncertainties.

\section{Acknowledgments}\label{sec:acknowledgments}
We thank the IT team of the INFN Rome section and the Physics Department at Sapienza University of Rome, and in particular Francesco Safai Tehrani, for their valuable support in accessing the local GPU computing infrastructure. We are also grateful to Michael Kagan and Lukas Heinrich for inspiring discussions, and to the Munich Institute for Astro-, Particle and BioPhysics (MIAPbP) for their hospitality and the stimulating environment provided during the workshop “Differentiable and Probabilistic Programming for Fundamental Physics” in June 2023.

\newpage
\appendix
\section{Model Architecture}\label{app:model}

The differentiable end-to-end model is based on two stacked neural networks inspired by the Graph Attention Network (GAT). 
The initial graph is built as fully connected using the recorded hits as input nodes. 

The node embedding is computed via an MLP which maps the input features $\mathbf{x}_i$ of each node $i$ 
into a hidden representation of dimension 512:
\begin{equation*}
\mathbf{q}_i = \texttt{attention\_query}(\mathbf{x}_i) \in \mathbb{R}^{512} \,.
\end{equation*}
The raw attention logit is computed as an edge feature:
\begin{equation*}
\mathbf{e}_{ij} = \texttt{attention\_logit}(\mathbf{q}_i, \mathbf{q}_j) \in \mathbb{R}^{512}\,,
\end{equation*}
where, to increase the expressivity of the network, $\mathbf{e}_{ij}$ is a vector-valued attention score 
of dimension 512. 
A dimension-wise softmax normalization is applied over all incoming edges of each receiving node:
\begin{equation*}
\alpha^{(d)}_{ij} = \frac{\exp(e^{(d)}_{ij})}{\sum_{k \in \mathcal{N}(j)} \exp(e^{(d)}_{kj})}, 
\quad d = 1, \ldots, 512 \,,
\end{equation*}
where $\mathcal{N}(j)$ denotes the neighbors of node $j$. The updated node features are computed via 
element-wise multiplication and aggregation:
\begin{equation*}
\mathbf{h}_j' = \sum_{i \in \mathcal{N}(j)} \boldsymbol{\alpha}_{ij} \odot \mathbf{q}_i \,,
\end{equation*}
where $\odot$ denotes the Hadamard product and $\boldsymbol{\alpha}_{ij} = [\alpha^{(1)}_{ij}, \ldots, \alpha^{(512)}_{ij}]^\top$.

The second GAT layer uses the output $\mathbf{h}_j'$ from the first layer as input. 
The attention mechanism is identical to the first layer, but includes an additional node update function 
that compresses the 512-dimensional representation to a scalar via an MLP followed by sigmoid activation:
\begin{equation*}
\mathbf{h}_j'' = \sigma\left(\texttt{node\_update}\left(\sum_{i \in \mathcal{N}(j)} \boldsymbol{\alpha}_{ij} \odot \mathbf{h}_i'\right)\right) \in [0,1] \,,
\end{equation*}
where $\sigma$ is the sigmoid function. Table~\ref{tab:architecture} summarizes the implementation details 
of both GAT layers. 

\begin{table}[h!]
\centering
\begin{tabular}{ll}
\hline
\textbf{Stage} & \textbf{Description} \\
\hline
Input graph & Nodes: $(y, z)$ coordinates of hits \\
& Edges: fully connected graph with self-edges  \\
\hline
GAT layer 1 & Attention query: MLP([4, 8, 16, 32, 64, 128, 256, 512]) \\
& Attention logit: MLP([1024, 512] $\rightarrow$ [512]) + LeakyReLU \\
\hline
GAT layer 2 & Attention query: Identity \\
& Attention logit: MLP([1024, 512] $\rightarrow$ [512]) + LeakyReLU \\
& Node update: MLP([512, 256, 128, 64, 32, 16, 8, 4, 2, 1]) + Sigmoid \\
\hline
\end{tabular}
\caption{Details of the implementation of the GAT layers used for the end-to-end architecture for the task of muon tracking.}
\label{tab:architecture}
\end{table}

\end{document}